\documentclass[aps,pra,floatfix,twocolumn,english,superscriptaddress]{revtex4}
\usepackage[latin1]{inputenc}
\usepackage{graphicx}
\usepackage{longtable}
\makeatletter

\usepackage{babel}
\makeatother
\begin{document}

\title{Theory of Cross Phase Modulation for the Vibrational Modes of Trapped Ions}

\author{X. Rebecca Nie}
\email{rebecca.nie@utoronto.ca}
\affiliation{Department of Physics and Centre of Quantum Information and Quantum
Control\\
University of Toronto, 60 St. George Street, Toronto, ON,  M5S 1A7, CANADA}
\author{Christian F. Roos}
\email{christian.roos@uibk.ac.at}
\affiliation{%
Institut f\"ur Experimentalphysik, Universit\"{a}t Innsbruck,
Technikerstra{\ss}e 25, A--6020
Innsbruck, Austria\\
}%
\affiliation{%
Institut f\"ur Quantenoptik und Quanteninformation der
\"Osterreichischen Akademie der Wissenschaften,
Technikerstra{\ss}e 21a, A--6020 Innsbruck, Austria\\
}%
\author{Daniel F. V. James}
\email{dfvj@physics.utoronto.ca}
\affiliation{Department of Physics and Centre of Quantum Information and Quantum Control\\
University of Toronto, 60 St. George Street, Toronto, ON,  M5S 1A7, CANADA}

\begin{abstract}
We analyze nonlinear coupling between individual vibrational quanta for trapped ions.  The nonlinear Coulomb interaction causes a Kerr-type Hamiltonian, for which we derive an analytical expression for the coupling constant $\chi$.  In contrast to a previously published formula \cite{Roos}, our result is in close agreement with experimental data.
\end{abstract}

\maketitle

It is well known that the nonlinear interaction at single quantum level is crucial to the development of quantum technologies.
Systems in which two field modes are coupled by a Kerr nonlinearity (ie. with a Hamiltonian $\hat{H}_{kerr}=\hbar\chi\hat{n}_r\hat{n}_s$) enable quantum nondemolition measurements \protect{\cite{Imoto}}, allow quantum gate operations in photonic quantum
computation \protect{\cite{Chuang},\cite{Munro}}, and assist nondestructive Bell-state detection \protect{\cite{Barrett}}.
However, it is often difficult to obtain the strong nonlinearities required to observe these
effects at the single quantum level \protect{\cite{Turchette},\cite{Schmidt}}.  One exception is provided by the joint vibrational modes of trapped ions, which have an intrinsic nonlinearity considerably stronger than that of photons, and can be utilized to our advantage \protect{\cite{Steane}}. In this paper, we shall present an analytical expression to describe this phonon-phonon interaction for single quanta and compare it to the experimental results.

Let us concentrate on the dispersive cross-Kerr effects causing shifts of the normal mode frequencies \protect{\cite{Odom}} rather than the resonant mode-mode coupling \protect{\cite{Marquet}}.
Suppose we have two equally charged ions of the same mass, confined in a three dimensional harmonic trapping potential.
We assume the strength of the trapping potential is the same in two transverse
directions, and is characterized by an angular frequency $\omega_{\perp},$ while in the axial direction, the trapping potential characterized by angular frequency $\omega_z$ is considerably weaker (Figure 1).  We have the following
classical expression for the system's kinetic energy and potential.
\begin{eqnarray}
\mathcal{T} &=& \frac{m}{2}(\dot{x_1}^2 + \dot{y_1}^2 + \dot{z_1}^2 + \dot{x_2}^2 + \dot{y_2}^2 + \dot{z_2}^2) \nonumber \\
&=& m(\dot{X}^2 +\dot{Y}^2 + \dot{Z}^2 + \dot{x}^2 + \dot{y}^2 + \dot{z}^2),\nonumber\\
\end{eqnarray}
and
\begin{eqnarray}
\mathcal{V} &=& \omega_\bot^2 \frac{m}{2}(x_1^2 + y_1^2 + x_2^2 + y_2^2) + \omega_z^2 \frac{m}{2}(z_1^2 + z_2^2) \nonumber \\
& &+\frac{q^2}{4 \pi \epsilon_0}\frac{1}{\sqrt{(x_1-x_2)^2 + (y_1 - y_2)^2 + (z_1-z_2)^2}} \nonumber\\
&=& \omega_\bot^2m(X^2 + Y^2) + \omega_z^2mZ^2\nonumber\\
&&+\omega_\bot^2m(x^2 + y^2 ) + \omega_z^2m z^2
+ \frac{q^2}{8 \pi \epsilon_0}\frac{1}{\sqrt{x^2 + y^2 + z^2}},\nonumber\\
\end{eqnarray}
where the center of mass and the relative motions are written as $X = \frac{1}{2}(x_1 + x_2)$, $Y = \frac{1}{2}(y_1 + y_2)$,
$Z = \frac{1}{2}(z_1 + z_2)$, and $x = \frac{1}{2}(x_1 - x_2)$, $y = \frac{1}{2}(y_1 - y_2)$, $z = \frac{1}{2}(z_1 - z_2)$ respectively.  Further, $m$ is the mass of one of the ions,
$q$ is the charge of one of the ions (for singly ionized ions $q=e$, the positively valued fundamental charge), and $\epsilon_0$ is the permittivity of free space.

\begin{figure}
\includegraphics[width=1.0 \columnwidth]{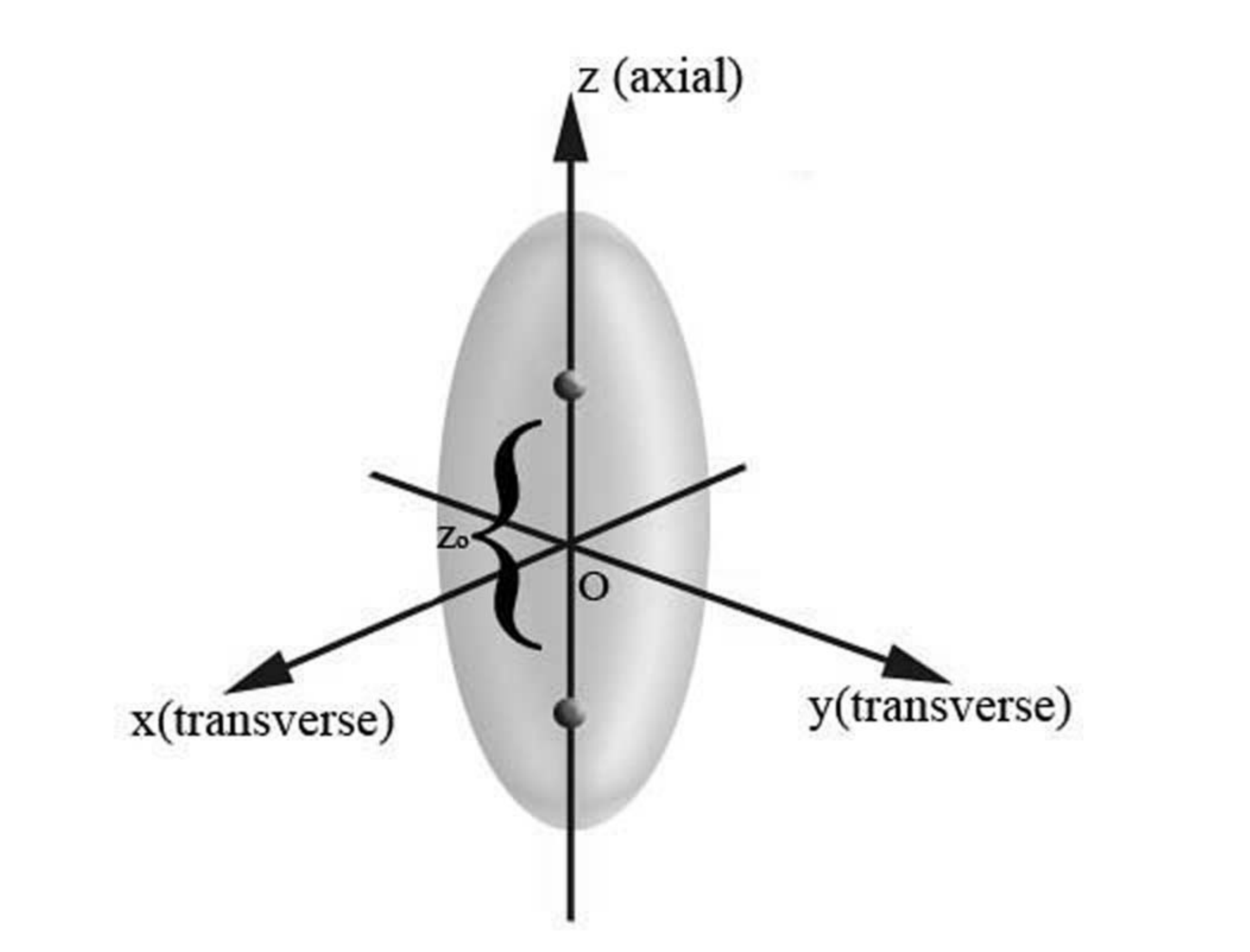}
\caption{Schematic diagram of the system. }
\end{figure}

Since the center of mass motion, characterized by $\{X,Y,Z\}$ is entirely decoupled from the relative motional degrees of freedom $\{x,y,z\}$, we ignore it;
this leaves us the following Lagrangian for the relative motion:
\begin{eqnarray}
\mathcal{L}_{rel} &=& \mathcal{T}_{rel} - \mathcal{V}_{rel} \nonumber\\
&= & m(\dot{x}^2 + \dot{y}^2 + \dot{z}^2) - \Big[m\omega_\bot^2(x^2 + y^2) + m\omega_z^2z^2\nonumber \\
& &+ \frac{q^2}{8\pi \epsilon_0}\frac{1}{\sqrt{x^2 + y^2 + z^2}}\Big], \nonumber
\end{eqnarray}
Defining the canonical momenta in the standard way (i.e. $p_x=\partial \mathcal{L}_{rel} /\partial \dot{x}$, and similarly for $p_y$ and $p_z$), this Lagrangian gives us the Hamiltonian for the relative motion
\begin{eqnarray}
\protect{\label{eqn1}}
\mathcal{H}_{rel} &=& \frac{1}{4m}(p_x^2 + p_y^2 + p_z^2) + m\omega_\bot^2(x^2 + y^2) + m\omega_z^2z^2 \nonumber \\
& &+ \frac{q^2}{8\pi\epsilon_0}\frac{1}{\sqrt{x^2 + y^2 + z^2}}.
\end{eqnarray}
Note that the {\em effective mass} for the relative motion is $2 m$, hence the cancellation of the factor $2$ in the denominator in the potential energy, and the appearance of $4$ in the denominator of the kinetic energy.  Rather than confuse things by introducing another symbol for this effective mass, we retain $m$, which is the mass of {\em one} of the the two ions.
The equilibrium separation of the ions is determined in the usual way by solving
$\partial\mathcal{V}_{rel}/\partial x=$$\partial\mathcal{V}_{rel}/\partial y=$$\partial\mathcal{V}_{rel}/\partial z=0$.
%
%
Provided $\omega_{\perp} >\omega_{z}$,  the
equilibrium separation of the ions is $x=0, y=0, z=z_0$, where
\begin{equation}
z_0=\sqrt[3]{\frac{q^2}{16 \pi \epsilon_0 m \omega_{z}^2}} = \sqrt[3]{\frac{\alpha\hbar c}{4m\omega_z^2}}.
\end{equation}
Here $\alpha$ is the fine structure constant (and we have assumed the ions are singly ionized, i.e. $q=e$), $\hbar$ is Planck's constant, and $c$ the speed of light.
We shall write $z = u + z_0$, $u$ being a new variable indicating the ions' displacement from their equilibrium separation.

It is insightful to expand the potential $\mathcal{V}_{rel}$ around the equilibrium position $(0,0,z_0)$.  First of all, let us write the potential
in terms of an expansion constant  $\varsigma$:
\begin{eqnarray}
\protect{\label{eqn11}}
&&\mathcal{V}[(0,0,z_0)+ \varsigma(x,y,u)]=
3m\omega_z^2 z_0^2 \nonumber \\
&&\,\,\,\,\,\,\,\,\,\,\,\, +\varsigma^2[m(\omega_\perp^2-\omega_z^2)(x^2 + y^2) + 3m\omega_z^2u^2] \nonumber \\
&&\,\,\,\,\,\,\,\,\,\,\,\, + \varsigma^3 \frac{m\omega_z^2}{z_0}[3(x^2+y^2)u - 2 u^3] \nonumber \\
&&\,\,\,\,\,\,\,\,\,\,\,\, + \varsigma^4\frac{3m\omega_z^2}{z_0^2}\Big[\frac{(x^2 + y^2)^2}{4} + \frac{2u^4}{3} - 2(x^2 + y^2)u^2\Big]. \nonumber
\\
\end{eqnarray}
The first term on the right hand side, which is independent of $\varsigma$, represents the constant potential energy of the equilibrium, and has no effect on the dynamics.  The terms $O(\varsigma^2)$ are harmonic potentials: immediately, we see from this term that the natural angular frequency of the ``rocking mode'' oscillations in the $x,y$-directions is $\omega_r = \sqrt{\omega_\perp^2 - \omega_z^2}$, while the ``stretch mode'' oscillation in the $z$ direction has the angular frequency $\omega_s = \sqrt{3}\omega_z$.
Keeping terms of $O(\varsigma^4)$ and setting the expansion constant  $\varsigma=1$, the Hamiltonian becomes
\begin{eqnarray}
\protect{\label{eqn2}}
\mathcal{H}_{rel} &=&
\mathcal{H}_0+\mathcal{V}^{(3)}+\mathcal{V}^{(4)}
\end{eqnarray}
where
\begin{eqnarray}
\mathcal{H}_0 &=& \frac{1}{4m}(p_x^2 + p_y^2 + p_z^2) + m\omega_r^2(x^2 + y^2) + m\omega_s^2u^2\nonumber\\
\end{eqnarray}
represents the harmonic motion (which, to a very good approximation, is the dominant feature of the ions' motion);
\begin{eqnarray}
\label{eqn3}
\mathcal{V}^{(3)} = \frac{m\omega_s^2}{z_0}[(x^2 + y^2)u - \frac{2}{3}u^3],
\end{eqnarray}
and
\begin{eqnarray}
\protect{\label{eqn4}}
\mathcal{V}^{(4)} = \frac{m\omega_s^2}{z_0^2}[\frac{1}{4}(x^2 + y^2)^2 + \frac{2}{3}u^4 - 2(x^2 + y^2)u^2],
\end{eqnarray}
are two terms which represent the lowest order perturbations of the harmonic terms.

To quantize the motion, we introduce the following operators:
\[\hat{x} = \sqrt{\frac{\hbar}{4m\omega_r}}(\hat{a} + \hat{a}^\dag),\qquad \hat{p_x} = i\sqrt{m\hbar\omega_r}(\hat{a} - \hat{a}^\dag); \]
\[\hat{y} = \sqrt{\frac{\hbar}{4m\omega_r}}(\hat{b} + \hat{b}^\dag),\qquad \hat{p_y} = i\sqrt{m\hbar\omega_r}(\hat{b} - \hat{b}^\dag); \]
\[\hat{u} = \sqrt{\frac{\hbar}{4m\omega_s}}(\hat{c} + \hat{c}^\dag),\qquad \hat{p_z} = i\sqrt{m\hbar\omega_s}(\hat{c} - \hat{c}^\dag). \]
The quantized
Hamiltonian is \begin{eqnarray}\hat{H} = \hat{H}_0 + \hat{H}_I,\nonumber\end{eqnarray} where
\begin{eqnarray}\hat{H}_0 = \hbar\omega_r(\hat{n}_x + \hat{n}_y +1) + \hbar\omega_s(\hat{n}_s + \frac{1}{2}), \nonumber
\end{eqnarray} and
\begin{eqnarray}\hat{H}_I = \hat{V}^{(3)} + \hat{V}^{(4)}.\nonumber\end{eqnarray}  $\hat{n}_x,$ $\hat{n}_y$, and $\hat{n}_s$ are the phonon number operators for their respective normal modes.

Now we are in a position to treat $\hat{H}_I$ with the standard second-order perturbation theory, but before we begin, recall that our goal is to extract the $\chi$ term from the Kerr-type Hamiltonian $
\hat{H}_{kerr} = \hbar\chi\hat{n}_r\hat{n}_s.$  To do that, we would like to calculate the shift in the stretch mode frequency as an effect of the coupling, when one phonon is inserted into the stretch mode. In other
words, we want
\begin{eqnarray}
\protect{\label{eqn5}}\delta\omega_s = \frac{\epsilon(n_s + 1, n_r^x, n_r^y) - \epsilon(n_s,n_r^x,n_r^y)}{\hbar}.
\end{eqnarray}
Let us start with $\hat{V}^{(4)}$, given by  eq.(\protect{\ref{eqn4}}), which, when quantized, becomes
\begin{eqnarray}
\protect{\label{eqn6}}
\hat{V}^{(4)} &=& -2\frac{m\omega_s^2}{z_0^2}(\hat{x}^2 + \hat{y}^2)\hat{u}^2 + \frac{2m\omega_s^2}{3z_0^2}\hat{u}^4 \nonumber \\
&=&-\hbar \omega_s \xi \Big(\frac{\omega_z}{\omega_r}\Big)\Big[ (\hat{n}_s + \frac{1}{2})(\hat{n}_x + \hat{n}_y + 1) \nonumber \\
& &\qquad + \frac{1}{2}(\hat{c}^2 + \hat{c}^{\dag2})(\hat{n}_x + \hat{n}_y + 1) \nonumber \\
& &\qquad  - \frac{1}{3}\frac{\omega_r}{\omega_s}(\hat{n_s}^2  + \hat{n}_s + \frac{1}{4})\Big] \nonumber \\
& & \qquad + \textrm{off resonance terms}.
\end{eqnarray}
Here $\xi = (2\hbar\omega_z/\alpha^2mc^2)^{1/3} $ is a dimensionless term of the order of $10^{-5}$ for atomic ions.
The off resonance terms, as well as the second term in Eq.(\protect{\ref{eqn6}}) will not
have any effect  to first order on the energy. Further, we shall neglect the third term, as it involves anharmonicity of the stretch mode rather than cross-coupling between the stretch and rocking motion.  Thus we find that $\hat{V}^{(4)}$
has the following contribution to the energy shift:
\begin{eqnarray}\protect{\label{eqn7}}&&\epsilon^{(4)}(n_s,n_r^x,n_r^y) \nonumber \\ &=& -\hbar\omega_s\xi\Big(\frac{\omega_z}{\omega_r}\Big)(n_s + \frac{1}{2})(n_r^x+n_r^y+1).\end{eqnarray}

Now we deal with $\hat{V}^{(3)}$,  eq.(\protect{\ref{eqn3}}), which can be quantized as
\begin{eqnarray}
\protect{\label{eqn8}}
\hat{V}^{(3)} &=& \zeta\Big\{\big[2(\hat{n}_x + \hat{n}_y +1)\hat{c} + 2(\hat{n}_x + \hat{n}_y +1)\hat{c}^\dag \nonumber \\
& & + \hat{c}\hat{a}^2  + \hat{a}^{\dag 2}\hat{c}  + \hat{b}^2\hat{c} + \hat{b}^{\dag 2}\hat{c}+\hat{a}^2\hat{c}^\dag + \hat{a}^{\dag 2}\hat{c}^\dag + \hat{b}^2\hat{c}^\dag \nonumber \\
& & + \hat{b}^{\dag 2}\hat{c}^\dag\big]  -\frac{2\omega_r}{3\omega_s}\big[\hat{c}(2\hat{n}_s +1) + \hat{c}^\dag(2\hat{n}_s +1) + \hat{c}^3 \nonumber \\
& & + \hat{c}\hat{c}^{\dag 2} + \hat{c}^\dag \hat{c}^2  + \hat{c}^{\dag 3}\big]\Big\},
\end{eqnarray}
where $\zeta = \sqrt{\xi\hbar^2\omega_s^3\omega_z/32\omega_r^2}$ is a constant with units of energy.

Some facts about the perturbation $\hat{V}^{(3)}$ are now apparent. Firstly,  $\hat{V}^{(3)}$ represents the creation and annihilation of vibrational quanta, and does not yield any value change in {\em energy} to first order.  Therefore, to calculate energy shifts, we need to seek out the cross-coupled terms  from in its second order perturbation.  Secondly, since the terms involving $\hat{c}^3$ and $\hat{c}^{\dag 3}$ do not give rise to any cross-coupled terms, they can be safely neglected. Finally, only a few states out
of the infinite number of possible states can yield nonzero value during the second order perturbation of the rest of $\hat{V}^{(3)}$.  All those states, their corresponding matrix elements, and their respective
differences between the unperturbed Hamiltonian's eigenvalues are listed in \textsc{Table 1}.

\begin{longtable}{|l|p{4cm}|p{2cm}|}
\endhead
\hline $|\varphi_m \rangle$ & $\langle \varphi_n|\hat{V}^{(3)}|\varphi_m \rangle$ & $E^{(0)}_n-E^{(0)}_m$ \\
\hline $|n_r^x,n_r^y,n_s+1\rangle$& $2\zeta(n_s+1)^{\frac{1}{2}}[(n_r^x+n_r^y +1 ) -\frac{\omega_r}{\omega_s}(n_s+1)]$  &$-\hbar\omega_s$\\
\hline $|n_r^x,n_r^y,n_s-1\rangle$& $2\zeta n_s^{\frac{1}{2}}[(n_r^x+n_r^y +1 )-\frac{\omega_r}{\omega_s}n_s]$ & $\hbar\omega_s$ \\
\hline $|n_r^x +2, n_r^y,n_s+1\rangle$ & $\zeta[(n_s+1)(n_r^x+1)(n_r^x+2)]^{\frac{1}{2}}$ & $-\hbar\omega_s - 2\hbar\omega_r$\\
\hline $|n_r^x +2, n_r^y,n_s-1\rangle$ & $\zeta[n_s(n_r^x+1)(n_r^x+2)]^{\frac{1}{2}}$ & $\hbar\omega_s - 2\hbar\omega_r$\\
\hline $|n_r^x -2, n_r^y,n_s+1\rangle$ & $\zeta[(n_s+1)n_r^x(n_r^x+2)]^{\frac{1}{2}}$ & $-\hbar\omega_s + 2\hbar\omega_r$\\
\hline $|n_r^x -2, n_r^y,n_s-1\rangle$ & $\zeta[n_sn_r^x(n_r^x-1)]^{\frac{1}{2}}$ & $\hbar\omega_s + 2\hbar\omega_r$\\
\hline $|n_r^x, n_r^y+2,n_s+1\rangle$ & $\zeta[(n_s+1)(n_r^y+1)(n_r^y+2)]^{\frac{1}{2}}$ & $-\hbar\omega_s - 2\hbar\omega_r$\\
\hline $|n_r^x, n_r^y+2,n_s-1\rangle$ & $\zeta[n_s(n_r^y+1)(n_r^y+2)]^{\frac{1}{2}}$ & $\hbar\omega_s - 2\hbar\omega_r$\\
\hline $|n_r^x, n_r^y-2 ,n_s+1\rangle$ & $\zeta[(n_s+1)n_r^y(n_r^y-1)]^{\frac{1}{2}}$ & $-\hbar\omega_s + 2\hbar\omega_r$\\
\hline $|n_r^x, n_r^y-2,n_s-1\rangle$ & $\zeta[n_sn_r^y(n_r^x-1)]^{\frac{1}{2}}$ & $\hbar\omega_s + 2\hbar\omega_r$\\
\hline
\caption*{\textsc{Table 1}: The states $|\varphi_m \rangle \equiv |m_r^x,m_r^y,m_s\rangle$ relevant to the second order perturbation of $\hat{V}^{(3)}$,
their corresponding matrix elements
($\langle \varphi_n|\hat{V}^{(3)}|\varphi_m \rangle \equiv
\langle n_r^x,n_r^y,n_s|\hat{V}^{(3)}|m_r^x,m_r^y,m_s \rangle$),
and their respective differences between the unperturbed Hamiltonian's  eigenvalues
($E^{(0)}_n-E^{(0)}_m\equiv E_{n_r^x,n_r^y,n_s}^{(0)} - E_{m_r^x,m_r^y,m_s}^{(0)}$
; the parameter $\zeta = \sqrt{\xi\hbar^2\omega_s^3\omega_z/32\omega_r^2}$ is a constant with units of energy.}
\end{longtable}

The next step is to sum these matrix elements in the standard manner given by second order perturbation theory , i.e.,
\begin{eqnarray}
\Delta E=
\sum_{\{m_r^x,m_r^y,m_s\}'}{\frac{|\langle n_r^x,n_r^y,n_s|\hat{V}^{(3)}|m_r^x,m_r^y,m_s\rangle|^2}{E_{n_r^x,n_r^y,n_s}^{(0)} - E_{m_r^x,m_r^y,m_s}^{(0)}}},\nonumber
\end{eqnarray}
where $\{m_r^x,m_r^y,m_s\}'$ denotes the condition $m_r^xm_r^ym_s\neq n_r^xn_r^yn_s$ \protect{\cite{Shankar}}.
Keeping only the cross-coupled terms, we find that the energy shift given by $\hat{V}^{(3)}$ is
\begin{eqnarray}
\protect{\label{eqn9}}
&&\epsilon^{(3)}(n_r^x,n_r^y,n_s) =\nonumber\\&&
-\hbar\omega_s\xi\Big(\frac{\omega_z}{\omega_r}\Big)
\Big[\frac{\omega_s^2}{8\omega_r^2-2\omega_s}n_s - \frac{1}{2}\Big(n_s+\frac{1}{2}\Big)\Big] \times \nonumber\\&&
\,\,\,\,\,\,\,\,\,\,\,\,\,\,\,\,\,\,\,\,\,\,\,\,\,\,\,\,\,\,\,\,\,\,\,\,\,\,\,\,\,\,\,\,\,\,\,\,\,\,\,\,\,\,\,\,\,\,\,\,\,\,\,\,\,\,\,\,\,\,\,\,(n_r^x + n_r^y + 1).
\end{eqnarray}
Combining Eq.(\protect{\ref{eqn7}}) and Eq.(\protect{\ref{eqn9}}), the energy shift is found to be
\begin{eqnarray}
\protect{\label{eqn14}}
&&\epsilon(n_r^x,n_r^y,n_s)= -\hbar\omega_s \Big(\frac{2\hbar\omega_z}{\alpha^2mc^2}\Big)^{\frac{1}{3}}
\Big(\frac{\omega_z}{\omega_r}\Big)\times\nonumber \\
&&\,\,\,\,\Big[\frac{1}{2}(n_s+\frac{1}{2}) + \frac{\omega_s^2}{8\omega_r^2-2\omega_s^2}n_s \Big] (n_r^x+n_r^y+1).
\end{eqnarray}
Higher order perturbations, obtained by keeping terms of higher order in $\varsigma$ in eq.(\ref{eqn11}), give rise to terms in higher powers of $\xi$, which, as mentioned above, is a small quantity for atomic ions.  Thus the contributions of these higher terms may safely be neglected. From Eq.(\protect{\ref{eqn14}}), we find the expression for the cross-Kerr coefficient $\chi$
\begin{eqnarray}
\protect{\label{eqn10}}
\chi &=& -\omega_s\Big(\frac{2\hbar\omega_z}{\alpha^2mc^2}\Big)^{\frac{1}{3}}
\Big(\frac{\omega_z}{\omega_r}\Big)\Big[\frac{1}{2} + \frac{\omega_s^2/2}{4\omega_r^2-\omega_s^2}\Big].
\end{eqnarray}
This formula differs from eq.(1) of ref.\cite{Roos} in that the first term in the square bracket is $1/2$ rather than $1$.  The distinction arises entirely from the more thorough treatment of the second-order shift in energies due to the $\hat{V}^{(3)}$ presented in this paper.

Our theoretical model can be readily compared to the experimental data presented by Roos \textsl{et al.}, where two $^{40}\rm{Ca}^+$ ions are confined in a linear Paul trap \protect{\cite{Roos}}. Their experimental results for the measured
cross-Kerr coupling constant as a function of $\omega_z$ are compared to Eq.(\protect{\ref{eqn10}}), and presented in Figure 2.
As we can see, the theoretical model and the experimental results are in good agreement, much better than the theoretical curve presented in ref.\cite{Roos}.


\begin{figure}
\includegraphics[width=1.0 \columnwidth]{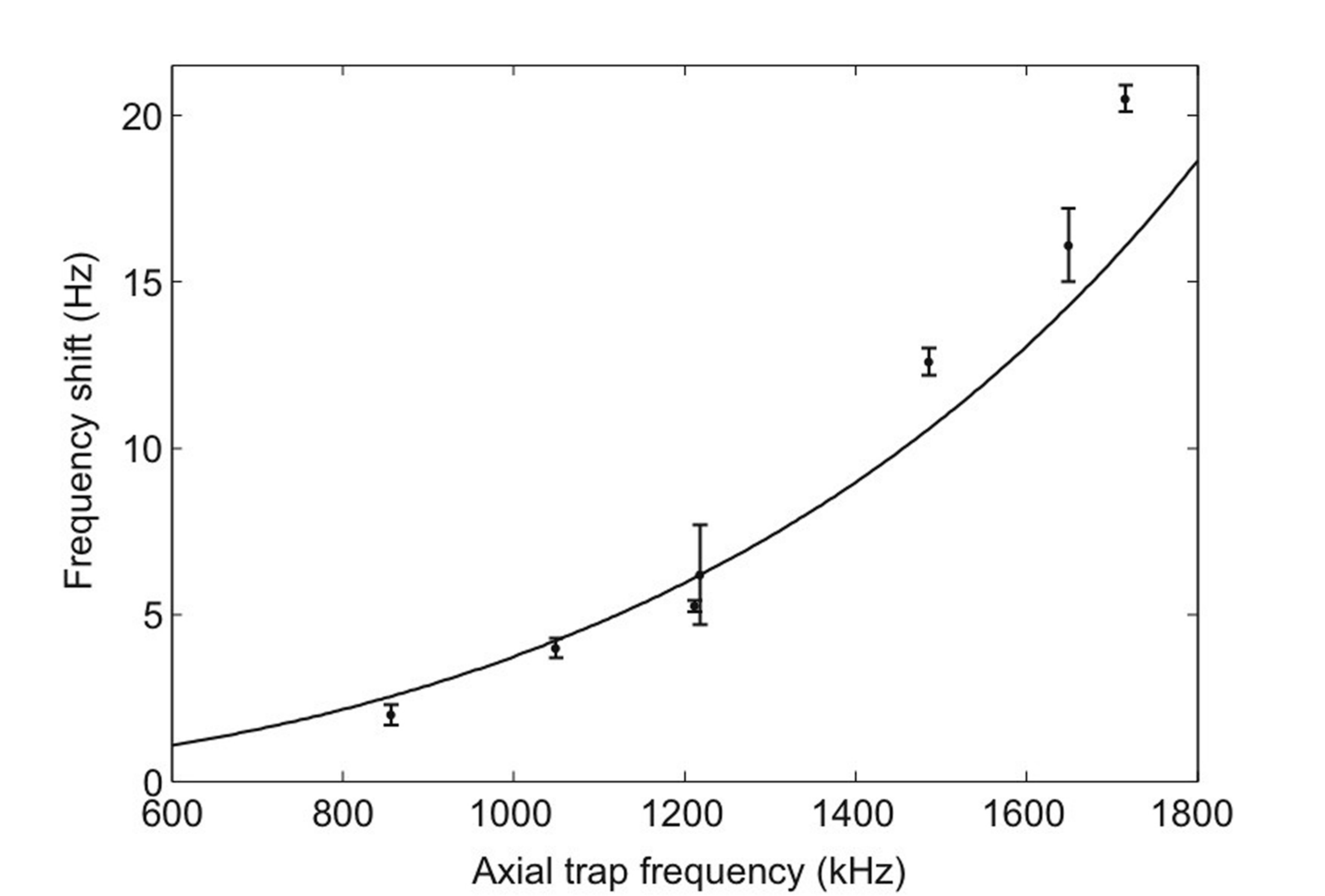}
\caption{Shift of the stretch mode frequency by a single phonon as a function of the axial trap frequency $\omega_z$. The solid line is the theoretical model predicted by Eq.(\protect{\ref{eqn10}}), and the points
are the data. }
\end{figure}

In conclusion, we have derived an analytical expression for the cross-Kerr coupling constant of two equally charged ions of the same mass trapped in a an-axial harmonic potential, while showing that the
theoretical model is verifiable via experimental means.  Since the analytical expression is reasonably general, it may describe other trapped ion systems giving rise to a cross coupling of harmonic oscillators
governed by a Kerr-like Hamiltonian $\hat{H}\propto\hat{n}_s\hat{n}_r.$  On the other hand, it would be interesting to further investigate the reason behind the experimental deviation from the theoretical model
in high $\omega_z$ regime.

We acknowledge financial support of NSERC.

\end{document}